\documentclass[journal,draftclsnofoot,onecolumn,12pt]{IEEEtran}

\usepackage{amsmath}
\usepackage{amssymb}
\usepackage[dvips]{graphicx}
\graphicspath{{./fig/}}
\usepackage{curves,color}
\usepackage{subfigure}
\usepackage{dsfont,slashbox,wasysym}
\usepackage{tikz}
\usepackage{pdfpages,enumerate}
\usepackage{multirow}
\usepackage{bm}
\usepackage{algorithmic}

\allowdisplaybreaks

\newtheorem{theorem}{Theorem}
\newtheorem{lemma}{Lemma}

\newtheorem{example}{Example}
\newtheorem{definition}{Definition}
\newtheorem{proposition}{Proposition}
\newtheorem{remark}{Remark}
\newtheorem{observation}{Observation}
\newtheorem{cor}{Corollary}

\RequirePackage[normalem]{ulem} 
\RequirePackage{color}\definecolor{RED}{rgb}{1,0,0}\definecolor{BLUE}{rgb}{0,0,1} 

\newenvironment{mydescription}[1]
{\begin{list}{}%
{\renewcommand\makelabel[1]{##1\hfill}%
\settowidth\labelwidth{\makelabel{#1}}%
\setlength\leftmargin{\labelwidth}
\addtolength\leftmargin{\labelsep}}}
{\end{list}}

\begin{document}
\title{Index Coding With Erroneous Side Information}
\author{Jae-Won~Kim~and~Jong-Seon~No, \IEEEmembership{Fellow,~IEEE}
\thanks{J.-W.~Kim and J.-S.~No are with the Department of Electrical and Computer Engineering, INMC, Seoul National University, Seoul 08826, Korea (e-mail: kjw702@ccl.snu.ac.kr, jsno@snu.ac.kr).}
}

\maketitle

\begin{abstract}
In this paper, new index coding problems are studied, where each receiver has erroneous side information. Although side information is a crucial part of index coding, the existence of erroneous side information has not yet been considered. We study an index code with receivers that have erroneous side information symbols in the error-free broadcast channel, which is called an index code with side information errors (ICSIE). The encoding and decoding procedures of the ICSIE are proposed, based on the syndrome decoding. Then, we derive the bounds on the optimal codelength of the proposed index code with erroneous side information. Furthermore, we introduce a special graph for the proposed index coding problem, called a $\delta_s$-cycle whose properties are similar to those of the cycle in the conventional index coding problem. Properties of the ICSIE are also discussed in the $\delta_s$-cycle and clique. Finally, the proposed ICSIE is generalized to an index code for the scenario having both additive channel errors and side information errors, called a generalized error correcting index code (GECIC). 
\end{abstract}

\vspace{10pt}
\section{Introduction}\label{sec:introduction}
\IEEEPARstart{I}{ndex} coding has attracted significant attention in various research areas since it was first introduced by Birk and Kol \cite{Informed}. Due to its relevance to various topics in information theory, lots of research has been done on index coding even though it was first considered for the satellite communication systems. Bar-Yossef {\em et~al.} proved that the optimal codelength of linear index codes is equal to the parameter $minrk$ of the fitting matrix of the side information graph and suggested an optimal construction method for the linear index codes \cite{ICSI}. In addition, it was proved that there is a nonlinear index code which outperforms the optimal linear index codes \cite{NIC}. 

It was proved in \cite{EQU} that every network coding problem can be converted to a corresponding index coding problem, and vice versa, for the given network structure. Furthermore, there was a trial relating the topological interference management (TIM) with index coding \cite{TIM}. Most index coding problems assume that the sender knows the side information graph and each receiver has some subsets of messages as side information. Recently, Kao {\em et~al.} researched a case where the sender only knows the probability distribution of side information in the receivers \cite{BIC}. In \cite{ICCSI}, Lee {\em et~al.} assumed that the sender knows the side information graph but a form of side information in each receiver is a linear combination of messages. While most of the index coding problems are studied in an error-free broadcast channel, there has been some work on the index codes with channel errors \cite{SECIC}-\cite{CCI}. In particular, Dau {\em et~al.} introduced error correcting index codes (ECIC) in the erroneous broadcast channel and algebraically analyzed them \cite{SECIC}. There has also been research on capacity analysis and application of index codes with side information over the additive white Gaussian noise (AWGN) channel \cite{ICG}, \cite{CCI}. As we can see from the previous works, index coding problems have been generalized further and become more realistic. 

   In this paper, new index coding problems with erroneous side information are presented, where each receiver has erroneous side information. In the conventional index coding, it is assumed that every receiver can exploit its side information directly because there is no side information error. However, there is always a possibility of memory errors in the receivers, which causes side information errors. Since there are some instances where side information is attained by sending messages through the broadcast channel, channel errors also cause erroneous side information. Thus, we have to consider a possibility having erroneous side information in the index coding problem.

   There are several applications of index coding with erroneous side information such as TIM. It is known that the interference channels of a receiver in TIM correspond to the messages, which are not the side information of the receiver in index coding. If the interference channels vary due to the moving receivers or are misunderstood as the non-interference channels from the false channel state information in TIM, this corresponds to index coding with erroneous side information. Thus, index coding with erroneous side information can be applied for the effective solutions of TIM.   

   In this paper, we propose the encoding and decoding procedures of index codes with side information errors (ICSIE), where each receiver has erroneous side information symbols in the error-free broadcast channel. One of the most important parameters in index coding is the codelength. Thus, the bounds on the optimal codelength of the proposed ICSIE are derived and its crucial graph, called a $\delta_s$-cycle, similar to the cycle in the conventional index coding is proposed. We relate the conventional index codes with the ICSIE by using the proposed bound and compare the cycle of the conventional index coding with the $\delta_s$-cycle of the proposed ICSIE. Furthermore, it is shown that there is a similarity between the generator matrix of the ICSIE and the transpose of the parity check matrix of the error correcting code when the corresponding side information graph is a clique. Finally, the ECIC in \cite{SECIC} is generalized by using the proposed ICSIE, which is called a generalized error correcting index code (GECIC). That is, we consider the more general scenario, where both channel errors and side information errors exist.
   
   The paper is organized as follows. The problem formulations for the ICSIE and the GECIC are given and the property of the generator matrix of the GECIC is derived in Section \ref{sec:preliminary}. In Section \ref{sec:Encoding}, the encoding and decoding procedures of the ICSIE are proposed. Then, the properties and bounds for the optimal codelength of the ICSIE are derived in Section \ref{sec:Properties}. Many properties of the ECIC in \cite{SECIC} are generalized to those of the GECIC by using the properties of the ICSIE in Section \ref{sec:Generalization}. Finally, conclusions are presented in Section \ref{Sec:Conclusion}.

\vspace{10pt}
\section{Problem Formulation and Some Results}\label{sec:preliminary}
\subsection{Notations}

Let $\mathbb{F}_q$ be the finite field of size $q$, where $q$ is a power of prime and $\mathbb{F}_q^{*}=\mathbb{F}_q\setminus\{0\}$. Let $Z[n]=\{1,2,\ldots,n\}$ for a positive integer $n$. For a vector $\bold{x}\in \mathbb{F}_q^n$, $\rm{\rm{wt}}(\bold{x})$ denotes the Hamming weight of $\bold{x}$. Let $\bold{x}_D$ be a sub-vector $(x_{i_1},x_{i_2},\ldots,x_{i_{|D|}})$ of a vector $\bold{x}=(x_1,x_2,\ldots,x_n)\in \mathbb{F}_q^n$ for a subset $D=\{i_1,i_2,\ldots,i_{|D|}\}\subseteq Z[n]$, where $i_1<i_2<\ldots<i_{|D|}$. We also introduce a sub-matrix $A_D$ of $A\in{\mathbb{F}_q}^{n\times N}$, that is, the matrix consisting of $|D|$ rows of $A$ as 
\begin{equation*}A_D=\begin{pmatrix}A_{i_1}\\A_{i_2}\\ \vdots\\  A_{i_{|D|}}\end{pmatrix}\end{equation*}
 where $A_i$ is the $i$th row of $A$. 

\begin{figure}[t]
\centering
\includegraphics[scale=0.5]{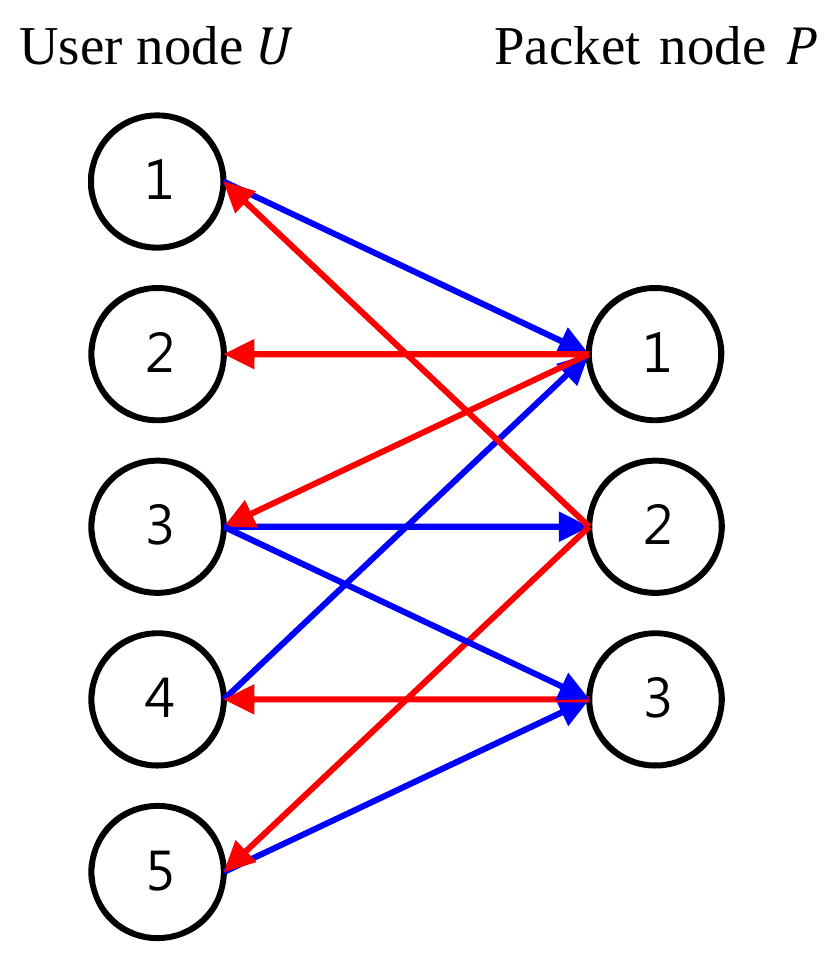}
\caption{An example of a directed bipartite side information graph with $n=3$ and $m=5$.} 
\label{fig:fig1}
\end{figure}

\subsection{Problem Formulation for Index Coding With Erroneous Side Information}

The conventional index coding is explained before introducing index coding with erroneous side information. We consider the index coding problem, where all information packets are elements in $\mathbb{F}_q$ and each receiver just wants one packet. This scenario is considered because any index coding problems can be converted to the problem of the above scenario if the size of packets is fixed. If the certain receiver wants $d$ packets, we can split the receiver into the $d$ receivers, each of which wants to receive one packet with the same side information. In this scenario, we can describe the conventional index coding with side information problem as follows. There are one sender which has $n$ information packets as $\bold{x}=(x_1,\ldots,x_n)\in\mathbb{F}_q^n$ and $m$ receivers (or users) $R_1,R_2,\ldots,R_m$, having sub-vectors of $\bold{x}$ as side information. Let ${{\mathcal{X}}_i}$ be the set of side information indices of the receiver $R_i$ for $i\in Z[m]$. That is, each receiver $R_i$ already knows the sub-vector $\bold{x}_{{\mathcal{X}}_i}$. Each receiver $R_i$ wants to receive one element in $\bold{x}$, called the wanted packet which is denoted by $x_{f(i)}$ and it is assumed that $\{f(i)\}\cap{{\mathcal{X}}_i}=\phi$. 

A side information graph shows the wanted packets and side information of all receivers. Fig. \ref{fig:fig1} shows the directed bipartite side information graph with five receivers and three packets. The edges from an user node to packet nodes represent the user's side information. For example, there are two edges from user 3 to packets 2 and 3 in Fig. \ref{fig:fig1}, which means that user 3 has packets 2 and 3 as side information. An edge from a packet node to an user node represents that the user wants this packet. For example, there is the edge from  packet 1 to user 2 in Fig. \ref{fig:fig1}, which means that user 2 wants to receive packet 1. Every packet node should have at least one outgoing edge and every user node should have one incoming edge. Otherwise, those nodes should be removed. If $m=n$ and $f(i)=i$ for all $i\in Z[m]$, we depict it as a unipartite side information graph \cite{ICSI}, which is given in the following section. It is assumed that the sender knows the side information graph $\mathcal{G}$ and broadcasts a codeword to receivers through the error-free channel in the conventional index coding problem. 

Having side information in each receiver is a crucial part of the index coding problem. However, a possibility to have erroneous side information has not been considered yet. Now, we propose a new index coding problem with erroneous side information by changing the side information condition in the conventional index coding problem, that is, each receiver $R_i$ has at most $\delta_s$ erroneous side information symbols. In the proposed index coding problem, a sender knows a side information graph $\mathcal{G}$ but does not know which side information is erroneous in each receiver. In addition, each receiver does not know which side information is erroneous. Furthermore, we can also consider additive channel errors in each receiver as in \cite{SECIC}. That is, each receiver receives $\bold{y}+\bm{\epsilon}_i$, where ${\bold y}$ is a codeword and $\bm{\epsilon}_i$ is an additive error vector such that $\bold{y},\bm{\epsilon}_i\in\mathbb{F}_q^N$ and ${\rm wt}(\bm{\epsilon}_i)\leq\delta_c$. 

\vspace{2mm}
\begin{definition}
A generalized error correcting index code with parameters $(\delta_s,\delta_c,\mathcal{G})$ over $\mathbb{F}_q$, denoted by a $(\delta_s,\delta_c,\mathcal{G})$-GECIC is a set of codewords having:
\vspace{2mm}
\begin{enumerate}
\item An encoding function $E: \mathbb{F}_q^n\rightarrow\mathbb{F}_q^N$.
\item A set of decoding functions $D_1,D_2,\ldots,D_m$ such that $D_i: {\mathbb{F}_q^N}\times{\mathbb{F}_q^{|{{\mathcal{X}}_i}|}}\rightarrow\mathbb{F}_q$ satisfying
\begin{equation}
D_i(E(\bold{x})+\bm{\epsilon}_i,\hat{\bold{x}}_{{\mathcal{X}}_i})=x_{f(i)}
\nonumber
\end{equation}
for all $i\in Z[m]$, $\bold{x}\in\mathbb{F}_q^n$, $\bm{\epsilon}_i\in\mathbb{F}_q^N$ with ${\rm wt}(\bm{\epsilon}_i)\leq\delta_c$, and ${\rm wt}(\bold{x}_{{\mathcal{X}}_i}-\hat{\bold{x}}_{{\mathcal{X}}_i})\leq\delta_s$, where $\hat{\bold{x}}_{{\mathcal{X}}_i}$ is the erroneous side information vector of the receiver $R_i$.  
\end{enumerate}
\vspace{2mm}
\end{definition}

Here, ${\rm wt}(\bold{x}_{{\mathcal{X}}_i}-\hat{\bold{x}}_{{\mathcal{X}}_i})\leq\delta_s$ means that the maximum number of side information errors is $\delta_s$ for each receiver. In this paper, we consider a linear index code. That is, $E(\bold{x})=\bold{x}G$ for all $\bold{x}\in\mathbb{F}_q^n$, where $G\in\mathbb{F}_q^{n\times N}$ is a generator matrix of the index code and $N$ denotes the codelength of the $(\delta_s,\delta_c,\mathcal{G})$-GECIC. Let $N_{\rm opt}^q(\delta_s,\delta_c,\mathcal{G})$ be the optimal codelength of the $(\delta_s,\delta_c,\mathcal{G})$-GECIC over $\mathbb{F}_q$. If $\delta_c=0$, a $(\delta_s,0,\mathcal{G})$-GECIC is called a $(\delta_s,\mathcal{G})$-index code with side information errors, denoted by a $(\delta_s,\mathcal{G})$-ICSIE and the optimal codelength $N_{\rm opt}^q(\delta_s,\delta_c,\mathcal{G})$ is modified to $N_{\rm opt}^q(\delta_s,\mathcal{G})$. Similarly, for $\delta_s=0$, we have $(\delta_c,\mathcal{G})$-ECIC and $N_{\rm opt}^q(\delta_c,\mathcal{G})$ as in \cite{SECIC}.

\subsection{Property of Generator Matrix of GECIC}

We find the property of the generator matrix for the proposed $(\delta_s,\delta_c,\mathcal{G})$-GECIC by generalizing Lemma 3.8 in \cite{SECIC}. Let $\mathcal{I}(q,\mathcal{G},\delta_s)$ be the set of vectors defined by
\begin{equation}
\mathcal{I}(q,\mathcal{G},\delta_s)=\bigcup_{i\in Z[m]}\mathcal{I}_i(q,\mathcal{G},\delta_s)
\nonumber
\end{equation}
where $\mathcal{I}_i(q,\mathcal{G},\delta_s)=\{\bold{z}\in\mathbb{F}_q^n: {\rm wt}(\bold{z}_{\mathcal{X}_i})\leq2\delta_s, z_{f(i)}\neq0\}$.
The support set of $\mathcal{I}(q,\mathcal{G},\delta_s)$ is defined as
\begin{equation}
J(\mathcal{G},\delta_s)=\bigcup_{i\in Z[m]}\big\{\{f(i)\}\cup Y_i\cup I_i: Y_i\subseteq\mathcal{Y}_i, I_i\subseteq\mathcal{X}_i\big\}
\nonumber
\end{equation}
where $|I_i|\leq2\delta_s$ and $\mathcal{Y}_i=Z[n]\setminus \big(\{f(i)\}\cup\mathcal{X}_i\big)$.

Then, the property of the generator matrix of the $(\delta_s,\delta_c,\mathcal{G})$-GECIC is given in the following theorem.
\vspace{2mm}
\begin{theorem}[Generalization of Lemma 3.8 in \cite{SECIC}]
A matrix $G$ is a generator matrix of the $(\delta_s,\delta_c,\mathcal{G})$-GECIC if and only if 
\begin{equation}
{\rm wt}(\bold{z}G)\geq2\delta_c+1 \textrm{, for all } \bold{z}\in\mathcal{I}(q,\mathcal{G},\delta_s).
\nonumber
\end{equation}
\begin{IEEEproof}
It will be proved in the similar manner as in \cite{SECIC}, that is, we use the similar concepts as Hamming spheres of the classical error correcting codes. Here, we find the set of message vectors which should be distinguished by the received codewords.
Let $B(\bold{x},\delta_c)$ be the Hamming sphere of the received codeword $\bold{y}$ defined by
\begin{equation}
B(\bold{x},\delta_c)=\{\bold{y}\in\mathbb{F}_q^N: \bold{y}=\bold{x}G+\bm{\epsilon}, \bm{\epsilon}\in\mathbb{F}_q^N, \textrm{and~}{\rm wt}(\bm{\epsilon})\leq\delta_c\}.
\nonumber
\end{equation}

First, we prove that the receiver $R_i$ can recover the wanted packet $x_{f(i)}$ if and only if 
\begin{equation}
B(\bold{x},\delta_c)\cap B(\bold{x}^{\prime},\delta_c)=\phi
\label{equ:B}
\end{equation}
for every pair $\bold{x}$ and $\bold{x}^{\prime}\in\mathbb{F}_q^n$ such that $x_{f(i)}\neq x^{\prime}_{f(i)}$ and ${\rm wt}(\bold{x}_{\mathcal{X}_i}-\bold{x}^{\prime}_{\mathcal{X}_i})\leq2\delta_s$.

The only difference from Lemma 3.8 in \cite{SECIC} is that we replace the condition $\bold{x}_{\mathcal{X}_i}=\bold{x}_{\mathcal{X}_i}^{\prime}$ with ${\rm wt}(\bold{x}_{\mathcal{X}_i}-\bold{x}_{\mathcal{X}_i}^{\prime})\leq2\delta_s$. The side information vectors $\bold{x}_{\mathcal{X}_i}$ and $\bold{x}_{\mathcal{X}_i}^{\prime}$ of the receiver $R_i$ satisfy the inequality ${\rm wt}(\bold{x}_{\mathcal{X}_i}-\bold{x}_{\mathcal{X}_i}^{\prime})\leq2\delta_s$ if they can be the same by changing at most $\delta_s$ side information symbols, respectively. Now, we prove the above statement as follows.

$\mathit{Sufficiency}$: $R_i$ has to recover $x_{f(i)}$ by using side information and the received codeword. If $x_{f(i)}\neq x_{f(i)}^{\prime}$ and $R_i$ cannot distinguish $\bold{x}$ and $\bold{x}^{\prime}$ from the side information, the received codewords of $\bold{x}$ and $\bold{x}^{\prime}$ should be distinguished, which corresponds to (\ref{equ:B}).

$\mathit{Necessity}$: If $B(\bold{x},\delta_c)\cap B(\bold{x}^{\prime},\delta_c)=\phi$ for every message pair $\bold{x}$ and $\bold{x}^{\prime}$ such that $x_{f(i)}\neq x_{f(i)}^{\prime}$ and ${\rm wt}(\bold{x}_{\mathcal{X}_i}-\bold{x}_{\mathcal{X}_i}^{\prime})\leq2\delta_s$, $R_i$ can distinguish all received codeword pairs of $\bold{x}$ and $\bold{x}^{\prime}$ and thus $x_{f(i)}$ can be recovered. For other cases, $R_i$ can distinguish $\bold{x}$ and $\bold{x}^{\prime}$ from the side information or $R_i$ does not need to distinguish $\bold{x}$ and $\bold{x}^{\prime}$.

Let $\bold{z}=\bold{x}-\bold{x}^{\prime}$. Since each receiver $R_i$ has to recover $x_{f(i)}$, (\ref{equ:B}) should be satisfied for all $i\in Z[m]$. That is, the matrix $G$ corresponds to the generator matrix of the $(\delta_s,\delta_c,\mathcal{G})$-GECIC if and only if ${\rm wt}(\bold{z}G)\geq2\delta_c+1$ for all $\bold{z}\in\mathcal{I}(q,\mathcal{G},\delta_s)$. 
\end{IEEEproof}
\label{theorem:Encoding}
\end{theorem}
\vspace{2mm}

Here are several remarks regarding the above theorem.
\vspace{2mm}
\begin{remark}
It is obvious that $G$ is a generator matrix of the $(\delta_s,\delta_c,\mathcal{G})$-GECIC if and only if 
\begin{equation}
{\rm wt}(\Sigma_{i\in K}z_i G_i)\geq2\delta_c+1
\nonumber
\end{equation}
for all $K\in J(\mathcal{G},\delta_s)$ and all $z_i\in\mathbb{F}_q^*$.
\end{remark}
\vspace{2mm}
\begin{remark}
Since the receiver $R_i$ is only interested in $x_{f(i)}$, it is possible to have $B(\bold{x},\delta_c)\cap B(\bold{x}^{\prime},\delta_c)\neq\phi$ for ${\rm wt}(\bold{x}_{\mathcal{X}_i}-\bold{x}_{\mathcal{X}_i}^{\prime})\leq2\delta_s$ and $x_{f(i)}=x_{f(i)}^{\prime}$. It means that $R_i$ does not need to distinguish $\bold{x}$ and $\bold{x}^{\prime}$ because $x_{f(i)}=x_{f(i)}^{\prime}$.
\end{remark}
\vspace{2mm}
\begin{remark}
For $(\delta_s,\mathcal{G})$-ICSIE, the inequality ${\rm wt}(\bold{z}G)\geq2\delta_c+1$ becomes $\bold{z}G\neq\bold{0}$.
If the side information is assumed to be erased, we have 
\begin{equation}
\mathcal{I}_i(q,\mathcal{G},\delta_s)=\{\bold{z}\in\mathbb{F}_{q}^n: {\rm wt}(\bold{z}_{\mathcal{X}_i})\leq\delta_s, z_{f(i)}\neq0\}.
\nonumber
\end{equation}
\end{remark}
\vspace{2mm}

We provide an example for the aforementioned theorem.
\vspace{2mm}
\begin{example}
Let $q=2$, $m=n=4$, $\delta_s=1$, $f(i)=i$, and $\mathcal{G}$ is the clique of size 4. In general, the uncoded case is the worst case in the error-free channel, that is, the codelength is $n=4$. However, we can construct a $(\delta_s=1,\mathcal{G})$-ICSIE with codelength $3$. From Theorem \ref{theorem:Encoding}, it is clear that $\mathcal{I}(q,\mathcal{G},\delta_s=1)$ includes all vectors in  $\mathbb{F}_2^4$ except $(1,1,1,1)$ and $(0,0,0,0)$. Assume that we have a $4\times3$ matrix $G$ as
\begin{equation}
G=\begin{pmatrix}1&0&0\\0&1&0\\0&0&1\\1&1&1\end{pmatrix}.
\nonumber
\end{equation}

Then, we have $\bold{z}G\neq\bold{0}$ for all $\bold{z}\in\mathcal{I}(q,\mathcal{G},\delta_s=1)$. Thus, the above matrix $G$ is a generator matrix of the $(\delta_s=1,\mathcal{G})$-ICSIE.
\label{ex: encoding}
\end{example}
\vspace{10pt}
\section{Encoding and Decoding of $(\delta_s,\mathcal{G})$-ICSIE}\label{sec:Encoding}

In this section, we propose the encoding and decoding procedures of the proposed index code with erroneous side information in the error-free broadcast channel, that is, the ICSIE. 

\subsection{Encoding Procedure}
\vspace{2mm}
In general, design of the index codes is to find a generator matrix with the minimum codelength for the given side information graph, called the optimal index codes. In fact, any linearly dependent equations of the message packets can be generated by their minimum set of linearly independent equations. Thus, design of the optimal $(\delta_s,\mathcal{G})$-ICSIE corresponds to finding the minimum number of linearly independent equations of message packets, whose generator matrix satisfies the property in Theorem \ref{theorem:Encoding} with no channel error. Thus, we have the following remark for the $(\delta_s,\mathcal{G})$-ICSIE. 

\vspace{2mm}
\begin{remark}
If a generator matrix $G$ of the $(\delta_s,\mathcal{G})$-ICSIE has rank less than or equal to the codelength $N$, the matrix deleting any dependent columns from $G$ can also be its generator matrix. Thus, the generator matrix $G_{\rm opt}$ of the optimal $(\delta_s,\mathcal{G})$-ICSIE should have the rank  $N_{\rm opt}^q(\delta_s,\mathcal{G})$.
\label{re:rank}
\end{remark}
\vspace{2mm}

We propose the optimal construction method of the $(\delta_s,\mathcal{G})$-ICSIE, which is similar to that of the conventional index code in \cite{ICSI}. First, we generalize two well known definitions, the fitting matrices and their minimum rank for the given side information graph in \cite{ICSI}.

\vspace{2mm}
\begin{definition}
There are $\binom {|\mathcal{X}_i|}{2\delta_s}$ ways to choose $2\delta_s$ elements of $\mathcal{X}_i$ for $i\in Z[m]$, where $\binom {|\mathcal{X}_i|}{2\delta_s}=1$ for $|\mathcal{X}_i|<2\delta_s$. Let $T_i=\{\bold{i}_1,\ldots,\bold{i}_{\binom {|\mathcal{X}_i|}{2\delta_s}}\}$ for $i\in Z[m]$, where $\bold{i}_{j}$ denotes the set of chosen indices from $\mathcal{X}_i$ with cardinality $2\delta_s$ for $|\mathcal{X}_i|\geq2\delta_s$ and otherwise, $T_i=\{\bold{i}_1\}=\{\mathcal{X}_i\}$. An $n\times\Sigma_{i\in Z[m]}\binom {|\mathcal{X}_i|}{2\delta_s}$ matrix $A_g$ is said to be a generalized fitting matrix for $\mathcal{G}$ if $A_g$ satisfies the followings:

\begin{enumerate}
\item $A_g=[A_{ab}^{(i)}]$ consists of $m$ disjoint $n\times\binom {|\mathcal{X}_i|}{2\delta_s}$ submatrices $A^{(i)}$ for $i\in Z[m]$.
\item For $i\in Z[m]$ and  $b\in Z\big[\binom {|\mathcal{X}_i|}{2\delta_s}\big]$, $A_{ab}^{(i)}=0$ for $a\in \bold{i}_b$ and $A_{ab}^{(i)}$ can take any value of $\mathbb{F}_q$ for $a\in\mathcal{X}_i\setminus\bold{i}_b$.
\item $A_{f(i)b}^{(i)}=1$ for $b\in Z\big[\binom {|\mathcal{X}_i|}{2\delta_s}\big]$ and $i\in Z[m]$.
\item $A_{ab}^{(i)}=0$ for $a\in\mathcal{Y}_i$, $b\in Z\big[\binom {|\mathcal{X}_i|}{2\delta_s}\big]$, and $i\in Z[m]$.
\end{enumerate}
\end{definition}

\vspace{2mm}
\begin{definition}
$minrk_q(\delta_s,\mathcal{G})=\min\{rk_q(A_g) : A_g$ fits $\mathcal{G}\}$, where $rk_q(A_g)$ denotes the rank of $A_g$ over $\mathbb{F}_q$.
\end{definition}
\vspace{2mm}

A matrix $A_g$ which fits $\mathcal{G}$ satisfies the property of the generator matrix of the $(\delta_s,\mathcal{G})$-ICSIE in Theorem \ref{theorem:Encoding} as follows.

\vspace{2mm}
\begin{lemma}
Let $A_g$ be a generalized fitting matrix for $\mathcal{G}$. Then, $A_g$ satisfies the property of the generator matrix for the $(\delta_s,\mathcal{G})$-ICSIE, that is, $\bold{z}A_g\neq\bold{0}$ for all $\bold{z}\in\mathcal{I}(q,\mathcal{G},\delta_s)$.
\begin{IEEEproof}
For $\bold{z}\in\mathcal{I}_i(q,\mathcal{G},\delta_s)$ and $A^{(i)}$, let $\mathcal{A}$ be a set of indices of nonzero elements in $\bold{z}_{\mathcal{X}_i}$ with $|\mathcal{A}|\leq2\delta_s$. Then, there is at least one component in $\bold{z}A^{(i)}$, whose value is equal to nonzero $z_{f(i)}$ because there is at least one column vector $u$ in $A^{(i)}$ such that $u_{f(i)}=1$, $u^{\intercal}_{\mathcal{A}}=\bold{0}$, and $u^{\intercal}_{\mathcal{Y}_i}=\bold{0}$. Since it is true for all $i\in Z[m]$, there is at least one nonzero component in $\bold{z}A_g$ for any vector $\bold{z}\in\mathcal{I}(q,\mathcal{G},\delta_s)$.
\end{IEEEproof}
\label{lemma:fitting}
\end{lemma}

\vspace{2mm}
Thus, $A_g$ can be a generator matrix of the $(\delta_s,\mathcal{G})$-ICSIE from Theorem \ref{theorem:Encoding}, but $A_g$ is not the optimal generator matrix of the $(\delta_s,\mathcal{G})$-ICSIE.
By using the generalized fitting matrices and their minimum rank, the optimal codelength of the proposed ICSIE can be given in the following theorem, which corresponds to generalization of Theorem 1 in \cite{ICSI} for the conventional index coding with $\delta_s=0$. 
\vspace{2mm}
\begin{theorem}[Generalization of Theorem 1 in \cite{ICSI}]
$N_{\rm opt}^q(\delta_s,\mathcal{G})=minrk_q(\delta_s,\mathcal{G})$.
\vspace{2mm}
\begin{IEEEproof}
From Lemma \ref{lemma:fitting}, any generalized fitting matrix $A_g$ for $\mathcal{G}$ can be a generator matrix of the $(\delta_s,\mathcal{G})$-ICSIE. From Remark \ref{re:rank}, an $n\times minrk_q(\delta_s,\mathcal{G})$ matrix which is made by deleting dependent column vectors of $A_g$ also satisfies the property of the generator matrix for the $(\delta_s,\mathcal{G})$-ICSIE. Thus, the existence of the $n\times minrk_q(\delta_s,\mathcal{G})$ generator matrix is proved. Next, we prove that $minrk_q(\delta_s,\mathcal{G})$ is a lower bound for codelength $N$. We can prove this by using a similar method as in \cite{ICSI}. Let $V=\{\bold{v}_1,\ldots,\bold{v}_N\}$ be a set of basis column vectors of a generator matrix of the $(\delta_s,\mathcal{G})$-ICSIE and $\bold{e}_j$ be the $j$th standard basis column vector. Let $W$ be the vector space spanned by $V\cup\{\bold{e}_j : j\in I(i)\}$, where $I(i)\subseteq\mathcal{X}_i$ and $|I(i)|=|\mathcal{X}_i|-2\delta_s$. 

Now, we prove that $\bold{e}_{f(i)}\in W$. Suppose that $\bold{e}_{f(i)}\notin W$. Then, there exists at least one vector $\bold{x}^{\intercal}\in W^\perp$ such that $x_{f(i)}\neq0$ and ${\rm wt}(\bold{x}_{\mathcal{X}_i})\leq2\delta_s$. If there is no such vector $\bold{x}^{\intercal}\in W^\perp$, all vectors in $W^{\perp}$ are perpendicular to $\bold{e}_{f(i)}$, which means that $\bold{e}_{f(i)}\in W$. Thus, if $\bold{e}_{f(i)}\notin W$, there exists at least one $\bold{x}^{\intercal}\in W^\perp$ such that $x_{f(i)}\neq0$ and ${\rm wt}(\bold{x}_{\mathcal{X}_i})\leq2\delta_s$. Clearly, $\bold{x}\in\mathcal{I}(q,\mathcal{G},\delta_s)$ and $\bold{x}$ is encoded to $\bold{0}$, which contradicts the property of the generator matrix. Thus, $\bold{e}_{f(i)}\in W$ for all $i\in Z[m]$. 

Since there are $\binom {|\mathcal{X}_i|}{2\delta_s}$ ways to make $I(i)$ for $i\in Z[m]$, there are $\binom {|\mathcal{X}_i|}{2\delta_s}$ ways to express $\bold{e}_{f(i)}=\Sigma_la_l\bold{v}_l+\Sigma_{k\in I(i)}b_k\bold{e}_k$, where $a_l$ and $b_k$ are elements of $\mathbb{F}_q$. That is, $\Sigma_la_l\bold{v}_l=\bold{e}_{f(i)}-\Sigma_{k\in I(i)}b_k\bold{e}_k$, which can be a column vector of $A^{(i)}$. Let $M$ be the matrix whose column vectors are the above column vectors $\Sigma_la_l\bold{v}_l=\bold{e}_{f(i)}-\Sigma_{k\in I(i)}b_k\bold{e}_k$ for all $I(i)$ and $i\in Z[m]$. Then, $M$ fits $\mathcal{G}$, which implies that any generator matrix of the $(\delta_s,\mathcal{G})$-ICSIE can be transformed into the generalized fitting matrix by linear column operations. Thus, $N\geq rk_q(M)\geq minrk_q(\delta_s,\mathcal{G})$.
\end{IEEEproof}
\label{theorem: rank encoding}
\end{theorem}
\vspace{2mm}

Thus, the generator matrix of the $(\delta_s,\mathcal{G})$-ICSIE can be constructed from the generalized fitting matrix for $\mathcal{G}$ by removing all dependent columns. There is an example for construction of a generator matrix for Example \ref{ex: encoding} as follows.
\vspace{2mm}
\begin{example}
Let $q=2$, $m=n=4$, $\delta_s=1$, $f(i)=i$, and $\mathcal{G}$ be the clique of size 4. From Theorem \ref{theorem: rank encoding}, $N_{\rm opt}^q(\delta_s,\mathcal{G})$ can be found by $minrk_q(\delta_s,\mathcal{G})$. A matrix $A_g$ which fits $\mathcal{G}$ is described as
\begin{equation}
A_g=\begin{pmatrix}A^{(1)}&A^{(2)}&A^{(3)}&A^{(4)}\end{pmatrix}
\nonumber
\end{equation} where
\begin{equation}
A^{(1)}=\begin{pmatrix}1&1&1\\0&0&*\\0&*&0\\{*}&0&0\end{pmatrix},  A^{(2)}=\begin{pmatrix}0&0&*\\1&1&1\\0&*&0\\{*}&0&0\end{pmatrix}
\nonumber
\end{equation}
\begin{equation}
A^{(3)}=\begin{pmatrix}0&0&*\\0&*&0\\1&1&1\\{*}&0&0\end{pmatrix},  A^{(4)}=\begin{pmatrix}0&0&*\\0&*&0\\{*}&0&0\\1&1&1\end{pmatrix}
\nonumber
\end{equation}
and $*$ denotes any value of $0$ or $1$. In order to minimize the rank of $A_g$, the value $0$ or $1$ is selected for $*$ in $A_g$ and the dependent columns are removed. Then, one of the optimal generator matrices $G_{\rm opt}$ is given as
\begin{equation}
G_{\rm opt}=\begin{pmatrix}1&0&0\\0&1&0\\0&0&1\\1&1&1\end{pmatrix}.
\nonumber
\end{equation}
\end{example}
\vspace{2mm}

\subsection{Decoding Procedure}
\vspace{2mm}
We propose the decoding procedure of the $(\delta_s,\mathcal{G})$-ICSIE similar to that of the ECIC in \cite{SECIC}. That is, we can consider the decoding procedure which is similar to the syndrome decoding of the classical linear error correcting code. First, we find the syndrome related to the side information errors, which is used to find the correct side information. This procedure is different from that of the ECIC because the side information errors exist in the proposed ICSIE. 

In order to introduce the decoding procedure, we assume the followings:
\begin{enumerate}
\item Each receiver receives a codeword $\bold{y}=\bold{x}G$ through the error-free channel.
\item The receiver $R_i$ has a side information vector $\hat{\bold{x}}_{\mathcal{X}_i}$ for $i\in Z[m]$, where the number of erroneous side information symbols is less than or equal to $\delta_s$. 
\item The receiver $R_i$ only wants to recover $x_{f(i)}$ for $i\in Z[m]$.
\end{enumerate}

In addition, we define the following notations:
\begin{enumerate}
\item $\tilde{\bold{x}}_{\delta_s}=\bold{x}_{\mathcal{X}_i}-\hat{\bold{x}}_{\mathcal{X}_i}$, where $\bold{x}_{\mathcal{X}_i}$ is a correct side information vector of $R_i$.
\item $H^{(i)}$ is a matrix orthogonal to span$(\{G_j\}_{j\in\{f(i)\}\cup\mathcal{Y}_i})$.
\item $H^{(i)}_e$ is a matrix orthogonal to span$(\{G_j\}_{j\in\mathcal{Y}_i})$ and not orthogonal to span$(\{G_j\}_{j\in\{f(i)\}})$. 
\end{enumerate}

\begin{figure}[t]
\noindent\fbox{\parbox{0.98\linewidth}{
\textbf{Algorithm 1: Decoding procedure for $R_i$} \\
\textbf{Input}: $\bold{y}$, $\hat{\bold{x}}_{\mathcal{X}_i}$, and $G$. \\
\textbf{Output}: $x_{f(i)}$. 
\begin{mydescription}{Step 1)}
\item[\textbf{Step 1)}] Compute the syndrome 
\begin{equation}
\bold{s}_i=H^{(i)}(\bold{y}-\hat{\bold{x}}_{\mathcal{X}_i}G_{\mathcal{X}_i})^{\intercal}=H^{(i)}(\tilde{\bold{x}}_{\delta_s}G_{\mathcal{X}_i})^{\intercal}.
\nonumber
\end{equation}
\item[\textbf{Step 2)}] Find one solution of $\bold{p}_i$ that satisfies $\bold{s}_i=H^{(i)}{\bold{p}_i}^{\intercal}$ under the condition that $\bold{p}_i$ is a linear combination of rows of $G_{\mathcal{X}_i}$, where the number of linearly combined rows in $G_{\mathcal{X}_i}$ is less than or equal to $\delta_s$.
\item[\textbf{Step 3)}]  Make the following equation
\begin{equation}
\tilde{\bold{y}}=\bold{y}-\hat{\bold{x}}_{\mathcal{X}_i}G_{\mathcal{X}_i}-\bold{p}_i=x_{f(i)}G_{f(i)}+(\bold{x}_{\mathcal{Y}_i}-\bold{b})G_{\mathcal{Y}_i}
\label{equ:decoding}
\end{equation}
where $\bold{b}\in\mathbb{F}_q^{|\mathcal{Y}_i|}$.
\item[\textbf{Step 4)}] Find $x_{f(i)}$ by multiplying the matrix ${H^{(i)}_e}^\intercal$ on both sides of (\ref{equ:decoding}).
\end{mydescription}
}}
\end{figure}

Then, the decoding procedure of the $(\delta_s,\mathcal{G})$-ICSIE for each receiver $R_i$ is described in Algorithm 1. We need the following theorem for $\bold{p}_i$ in (\ref{equ:decoding}) of the proposed decoding procedure.

\vspace{2mm}
\begin{theorem}
Let $\eta_i$ be a subset of $\mathcal{X}_i$ with $|\eta_i|\leq\delta_s$. Let $\bold{p}_i$ be a solution of $\bold{s}_i=H^{(i)}\bold{p}_i$, where $\bold{p}_i\in$ span$(\{G_j\}_{j\in\eta_i})$. Then, $\bold{p}_i$ is given as $\bold{p}_i=\tilde{\bold{x}}_{\delta_s}G_{\mathcal{X}_i}+\bold{k}$, where $\bold{k}\in$span$(\{G_j\}_{j\in\mathcal{Y}_i})$.
\label{theorem:decoding}
\end{theorem}
\vspace{2mm}
\begin{IEEEproof}
We can find a solution $\bold{p}_i$ for  $\bold{s}_i=H^{(i)}\bold{p}_i^{\intercal}$, under the condition that $\bold{p}_i$ is a linear combination of rows of $G_{\mathcal{X}_i}$, where the number of linearly combined rows in $G_{\mathcal{X}_i}$ is less than or equal to $\delta_s$ because there exist at least one such $\bold{p}_i$ due to $\tilde{\bold{x}}_{\delta_s}G_{\mathcal{X}_i}$. Moreover, if we find a solution $\bold{p}_i$ under the condition mentioned above, then we have
\begin{equation}
\bold{p}_i=\tilde{\bold{x}}_{\delta_s}G_{\mathcal{X}_i}+\bold{k}
\nonumber
\end{equation}
due to the property of the generator matrix. Specifically, from
\begin{equation}
\bold{s}_i=H^{(i)}(\tilde{\bold{x}}_{\delta_s}G_{\mathcal{X}_i})^{\intercal}=H^{(i)}\bold{p}_i^{\intercal}
\nonumber
\end{equation}
we have $H^{(i)}(\tilde{\bold{x}}_{\delta_s}G_{\mathcal{X}_i}-\bold{p}_i)^{\intercal}=0$. Thus, $\tilde{\bold{x}}_{\delta_s}G_{\mathcal{X}_i}-\bold{p}_i=aG_{f(i)}-\bold{b}G_{\mathcal{Y}_i}$ and 
\begin{equation}
aG_{f(i)}-\bold{b}G_{\mathcal{Y}_i}-\tilde{\bold{x}}_{\delta_s}G_{\mathcal{X}_i}+\bold{p}_i=\bold{0}
\label{equ:proof2}
\end{equation}
where $a\in\mathbb{F}_q$ and $\bold{b}\in\mathbb{F}_q^{|\mathcal{Y}_i|}$. Then, we can easily check that LHS of (\ref{equ:proof2}) is $\bold{x}G$ such that ${\rm wt}(\bold{x}_{\mathcal{X}_i})\leq2\delta_s$. Since RHS of  (\ref{equ:proof2}) is zero, $a$ should be zero by Theorem \ref{theorem:Encoding}. Therefore, $\bold{p}_i=\tilde{\bold{x}}_{\delta_s}G_{\mathcal{X}_i}+\bold{b}G_{\mathcal{Y}_i}$.
\end{IEEEproof}
\vspace{2mm}

Using Theorem \ref{theorem:decoding}, (\ref{equ:decoding}) can be given as $\tilde{\bold{y}}=x_{f(i)}G_{f(i)}+(\bold{x}_{\mathcal{Y}_i}-\bold{b})G_{\mathcal{Y}_i}$ and from Step 4, we have $\tilde{\bold{y}}{H^{(i)}_e}^{\intercal}=x_{f(i)}G_{f(i)}{H^{(i)}_e}^{\intercal}$. Thus, $x_{f(i)}$ can easily be obtained.

\vspace{2mm}
\begin{remark}
An interesting fact of this decoding procedure is that we can decode $x_{f(i)}$ even if we do not know the exact $\tilde{\bold{x}}_{\delta_s}G_{\mathcal{X}_i}$ as in the following example. 
\end{remark}
\vspace{2mm}
\begin{example}
Let $q=2, m=n=9, \delta_s=1$, and $f(i)=i$ for $i\in Z[9]$. Suppose that $\mathcal{X}_i=Z[9]\setminus\{i\}$ for $i\in Z[8]$ and $\mathcal{X}_9=\{2,3,5,6,7,8\}$. It is easy to check that one of the possible generator matrices of the above setting is given as
\begin{equation}
G=\begin{pmatrix}0&0&0&0&0&1\\0&0&0&0&1&0\\1&0&0&1&0&0\\0&0&1&0&0&0\\1&1&0&0&0&0\\0&1&1&1&0&0\\0&0&1&1&1&0\\0&0&0&1&1&1\\1&1&0&1&0&1\end{pmatrix}.
\nonumber
\end{equation}

For the message vector $\bold{x}=(1,1,1,1,0,0,0,0,1)$, we have the received codeword $\bold{y}=\bold{x}G=(0,1,1,0,1,0)$ in the error-free channel. In this case, we focus on the decoding procedure of the receiver $R_9$. We assume that $\hat{\bold{x}}_{\mathcal{X}_9}=(1,1,0,0,0,1)$. That is, the receiver $R_9$ has erroneous side information $\hat{x}_8$. 

The decoding procedure is described as follows:
\begin{enumerate}
\item We compute $\bold{y}-\hat{\bold{x}}_{\mathcal{X}_9}G_{\mathcal{X}_9}=(1,1,1,0,1,1)$.
\item Then, we can make $H^{(9)}$ from $G_{\{9\}\cup\mathcal{Y}_9}$ as
\begin{equation}
H^{(9)}=\begin{pmatrix}1&1&0&0&0&0\\0&1&0&1&0&0\\0&0&0&0&1&0\end{pmatrix}.
\nonumber
\end{equation}
\item Also, we can make $H^{(9)}_e$ as
\begin{equation}
H^{(9)}_e=\begin{pmatrix}1&0&0&0&0&0\\1&1&0&0&0&0\\0&1&0&1&0&0\\0&0&0&0&1&0\end{pmatrix}.
\nonumber
\end{equation}
\item Compute the syndrome as
\begin{equation}
\bold{s}_9=H^{(9)}(\bold{y}-\hat{\bold{x}}_{\mathcal{X}_9}G_{\mathcal{X}_9})^{\intercal}=\begin{pmatrix}0\\1\\1\end{pmatrix}.
\nonumber
\end{equation}
\item Find a solution $\bold{p}_{9}$ for $H^{(9)}\bold{p}_9^{\intercal}=\bold{s}_9$ under the decoding condition. 

Then, we have $\bold{p}_9=(0,0,0,1,1,1)$ and $(0,0,1,1,1,0)$. In fact, we need just one of two solutions. Choosing the first solution for $\bold{p}_9$ means that the receiver $R_9$ decides $\hat{x}_8$ as the erroneous side information while choosing the second solution means that $\hat{x}_7$ is decided as the erroneous side information. 
\item If $\hat{x}_8$ is chosen as the erroneous side information, then (\ref{equ:decoding}) in Algorithm 1 becomes 
\begin{equation}
\bold{y}-\hat{\bold{x}}_{\mathcal{X}_9}G_{\mathcal{X}_9}-(0,0,0,1,1,1)=(1,1,1,1,0,0)={x}_9G_9+(\bold{x}_{\mathcal{Y}_i}-\bold{b})G_{\mathcal{Y}_9}.
\nonumber
\end{equation}
Multiplying ${H^{(9)}_e}^\intercal$ on both sides leads to $(1,0,0,0)=x_9(1,0,0,0)$. Thus, $x_9=1$, which is the correct value. 
\item If $\hat{x}_7$ is chosen as the erroneous side information, then (\ref{equ:decoding}) in Algorithm 1  becomes 
\begin{equation}
\bold{y}-\hat{\bold{x}}_{\mathcal{X}_9}G_{\mathcal{X}_9}-(0,0,1,1,1,0)=(1,1,0,1,0,1)={x}_9G_9+(\bold{x}_{\mathcal{Y}_i}-\bold{b})G_{\mathcal{Y}_9}.
\nonumber
\end{equation}
Multiplying ${H^{(9)}_e}^\intercal$ on both sides leads to $(1,0,0,0)=x_9(1,0,0,0)$. Thus, $x_9=1$, which is also the correct value. 
\end{enumerate}
\label{ex:decoding}
\end{example}
\vspace{10pt}
\section{Properties and Bounds for Codelength of $(\delta_s,\mathcal{G})$-ICSIE}
\label{sec:Properties}

In this section, we introduce a new type of graphs, called a $\delta_s$-cycle for encoding of the index codes and derive some bounds for the optimal codelength of the $(\delta_s,\mathcal{G})$-ICSIE.

\subsection{$\delta_s$-cycle}
First, we define a $\delta_s$-cycle in $\mathcal{G}$ and generalize the generalized independent set and the generalized independence number of $\mathcal{G}$ in \cite{SECIC}. Let $\Phi$ be the set of subsets of $Z[n]$ defined by
\begin{equation}
\Phi=\{B\subseteq Z[n]\big||\mathcal{X}_i\cap B|\geq2\delta_s+1 \textrm{ for all } i\in Z[m] \textrm{ s.t. } f(i)\in B\}
\nonumber
\end{equation}
for a side information graph $\mathcal{G}$ of the $(\delta_s,\delta_c,\mathcal{G})$-GECIC. 

\vspace{2mm}
\begin{definition}
A subgraph $\mathcal{G}^{\prime}$ of $\mathcal{G}$ is called a $\delta_s$-cycle if the set of packet node indices of $\mathcal{G}^{\prime}$ is an element of $\Phi$ (say $B$) and the set of user node indices of $\mathcal{G}^{\prime}$ consists of $i\in Z[m]$ such that $f(i)\in B$ and its edges consist of the corresponding edges in $\mathcal{G}$. The graph $\mathcal{G}$ is said to be $\delta_s$-acyclic if there is no $\delta_s$-cycle in $\mathcal{G}$.
\label{def:delta cycle}
\end{definition} 

\vspace{2mm}
\begin{definition}
A set of packet node indices of a $\delta_s$-cycle is called a $\delta_s$-cycle induced set. Two $\delta_s$-cycles are said to be disjoint if their $\delta_s$-cycle induced sets are disjoint.
\end{definition}

\vspace{2mm}
\begin{definition}[Generalization of Definition 4.1 in \cite{SECIC}]
A subset $Q$ of $Z[n]$ is called a $\delta_s$-generalized independent set in $\mathcal{G}$ if every nonempty subset $K$ of $Q$ belongs to $J(\mathcal{G},\delta_s)$.
\end{definition}

\vspace{2mm}
\begin{definition}[Generalization of Definition 4.2 in \cite{SECIC}]   
Let $\gamma(\mathcal{G})$ be the largest size of a $\delta_s$-generalized independent set in $\mathcal{G}$, which is called the $\delta_s$-generalized independence number. 
\end{definition}

\vspace{2mm}
In the following lemma, we will show the relationship between a $\delta_s$-generalized independent set and a $\delta_s$-acyclic graph $\mathcal{G}$.

\vspace{2mm}
\begin{lemma}
Let $Q$ be a set of the packet node indices in $\mathcal{G}$. Then, $Q$ is a $\delta_s$-generalized independent set if and only if the side information graph $\mathcal{G}$ is $\delta_s$-acyclic.
\vspace{2mm}
\begin{IEEEproof}

$\mathit{Sufficiency}$: Let ${\mathcal{G}^{\prime}}$ be a subgraph of $\mathcal{G}$, where packet nodes of ${\mathcal{G}^{\prime}}$ is a subset of packet nodes of $\mathcal{G}$ and the sets of user nodes and edges of ${\mathcal{G}^{\prime}}$ are determined by $\mathcal{G}$. Suppose that $\mathcal{G}$ is $\delta_s$-cyclic. Then, we can make a subgraph $\mathcal{G}^{\prime}$ such that each receiver has side information symbols whose number is larger than or equal to $2\delta_s+1$. Let $Q^{\prime}\subseteq Q$ be the set of packet node indices of $\mathcal{G}^{\prime}$. Then, there is no vector, whose Hamming weight is $|Q^{\prime}|$ in $\mathcal{I}(q,\mathcal{G}^{\prime},\delta_s)$. Thus, $Q^{\prime}\notin J(\mathcal{G}^{\prime},\delta_s)$ and $Q^{\prime}$ is not also included in $J(\mathcal{G},\delta_s)$. It contradicts the assumption and thus $\mathcal{G}$ is $\delta_s$-acyclic.

$\mathit{Necessity}$: If $\mathcal{G}$ is $\delta_s$-acyclic, every subgraph $\mathcal{G}^{\prime}$ is also $\delta_s$-acyclic. That is, every nonempty subset $Q^{\prime}\subseteq Q$ belongs to $J(\mathcal{G},\delta_s)$ because there exists at least one receiver, whose number of side information symbols is less than or equal to $2\delta_s$.   
\end{IEEEproof}
\label{lemma:equivalence}
\end{lemma}

\vspace{2mm}

The important theorem for a $\delta_s$-cycle is given as follows.
\vspace{2mm}
\begin{theorem}
$\Phi=\phi$ if and only if $N_{\rm opt}^q(\delta_s,\mathcal{G})=n$ for the $(\delta_s,\mathcal{G})$-ICSIE.
\vspace{2mm}
\begin{IEEEproof}

$\mathit{Sufficiency}$: From Lemma \ref{lemma:equivalence}, there is an equivalence between a $\delta_s$-generalized independent set and a $\delta_s$-acyclic graph. If the set of packet node indices of $\mathcal{G}$ is the $\delta_s$-generalized independent set, $\mathcal{I}(q,\mathcal{G},\delta_s)$ is the set of all vectors in $\mathbb{F}_q^n$ except for the all zero vector. Thus, if $\Phi=\phi$, $Z[n]$ is the $\delta_s$-generalized independent set. Thus, all rows of a generator matrix of the $(\delta_s,\mathcal{G})$-ICSIE should be linearly independent from Theorem \ref{theorem:Encoding}. Then, we have $N_{\rm opt}^q(\delta_s,\mathcal{G})=n$.

$\mathit{Necessity}$: Suppose that $\Phi\neq\phi$. Then, we prove that the codelength can be reduced by at least one. We choose one $\delta_s$-cycle $\mathcal{G}^{\prime}$ in $\mathcal{G}$ and let $|B|$ be the number of packet nodes in $\mathcal{G}^{\prime}$. In this case, if we encode $\bold{x}^{\prime}\in\mathbb{F}^{|B|}_q$ as $(x_1^{\prime}+x_2^{\prime},x_2^{\prime}+x_3^{\prime},\ldots,x_{|B|-1}^{\prime}+x^{\prime}_{|B|})$ whose length is $|B|-1$, it satisfies the property of the generator matrix in Theorem \ref{theorem:Encoding} as follows.

Then, the generator matrix $G^{\prime}$ of this code is given as
\begin{equation}
G^{\prime}=\begin{pmatrix}1\\1&1\\&1&1\\&&1&\\&&&\ddots\\&&&&1\\&&&&1\end{pmatrix}.
\nonumber
\end{equation}  
We can easily find that any $|B|-1$ rows of $G^{\prime}$ are linearly independent and $|B|$ rows of $G^{\prime}$ are linearly dependent. Thus, if there is no vector whose Hamming weight is $|B|$ in $\mathcal{I}(q,\mathcal{G}^{\prime},\delta_s)$, $G^{\prime}$ satisfies the property of the generator matrix. Since $|\mathcal{X}^{\prime}_i|\geq2\delta_s+1$ for every user $i$ in $\mathcal{G}^{\prime}$ and $|B|\geq2\delta_s+2$, there is no vector whose Hamming weight is $|B|$ in $\mathcal{I}(q,\mathcal{G}^{\prime},\delta_s)$. Since there is the generator matrix $G^{\prime}$ whose codelength is $|B|-1$, $N_{\rm opt}^q(\delta_s,\mathcal{G})\neq n$. It contradicts the assumption and thus $\Phi=\phi$.
\end{IEEEproof}
\label{theorem:delta cycle}
\end{theorem}
\vspace{2mm}
\begin{remark}
From Theorem \ref{theorem:delta cycle}, we can conclude that a $\delta_s$-cycle is very crucial for index coding with erroneous side information because the existence of a $\delta_s$-cycle is a necessary and sufficient condition for a possibility to reduce the codelength of the $(\delta_s,\mathcal{G})$-ICSIE.
\end{remark}
\vspace{2mm}

Next, we compare the $\delta_s$-cycle of the ICSIE with the cycle of the conventional index code in the following proposition, where we can find similarities between the $\delta_s$-cycle and the cycle.
\vspace{2mm}
\begin{proposition}
The $\delta_s$-cycle of the ICSIE has very similar properties as those of cycles in the conventional index coding problem.
\vspace{2mm}

The cycle properties of the conventional index coding problem in \cite{DIC} are given as:
\begin{enumerate}
\item (Theorem 1 in \cite{DIC}) If $\mathcal{G}$ is acyclic, $N_{\rm opt}^q(0,\mathcal{G})=n$.
\item (Lemma 2 in \cite{DIC}) If a packet node has one outgoing edge, we say that the packet node is a unicast packet node. That is, there is only one user who wants this packet. We can reduce the codelength by the number of disjoint cycles which only consist of unicast packet nodes. 
\item (Lemma 3 in \cite{DIC}) If all distinct cycles of $\mathcal{G}$ are edge-disjoint and involve only unicast packets, then $N_{\rm opt}^q(0,\mathcal{G})=n-C(\mathcal{G})$, where $C(\mathcal{G})$ is defined as the largest number of edge-disjoint cycles in $\mathcal{G}$ that involve only unicast packets.
\end{enumerate}

We compare properties of the $\delta_s$-cycle in the ICSIE with those of the cycle in the conventional index coding problem \cite{DIC}. If $\delta_s=0$, there is at least one cycle in a $0$-cycle because every receiver node has at least one outgoing edge. Then, from Theorem \ref{theorem:delta cycle}, the corresponding properties of the $\delta_s$-cycle of the $(\delta_s,\mathcal{G})$-ICSIE are given as follows:
\begin{enumerate}
\item $\mathcal{G}$ is $\delta_s$-acyclic if and only if $N_{\rm opt}^q(\delta_s,\mathcal{G})=n$.
\item Let $\beta$ be the maximum number of disjoint $\delta_s$-cycles in $\mathcal{G}$. Then, the optimal codelength can be reduced by the number of disjoint $\delta_s$-cycles, that is, $N_{\rm opt}^q(\delta_s,\mathcal{G})\leq n-\beta$. This is given in the proof of the necessity of Theorem \ref{theorem:delta cycle}.
\item The last property is given in the corollary below.
\end{enumerate}
\label{prop:delta cycle}
\end{proposition}
\vspace{2mm}

Note that the cycle property 1) of the conventional index coding is the sufficient condition but the corresponding property of the ICSIE is the necessary and sufficient condition.

\vspace{2mm}
\begin{cor} 
Suppose that there is no $\delta_s$-cycle in a subgraph $\mathcal{G}^{\prime}$ constructed by removing one packet node from each of $\beta$ $\delta_s$-cycles and the corresponding edges. Then, $N_{\rm opt}^q(\delta_s,\mathcal{G})=n-\beta$.
\label{cor:delta cycle}

\begin{figure}[t]
\centering
\subfigure[]{\includegraphics[scale=0.5]{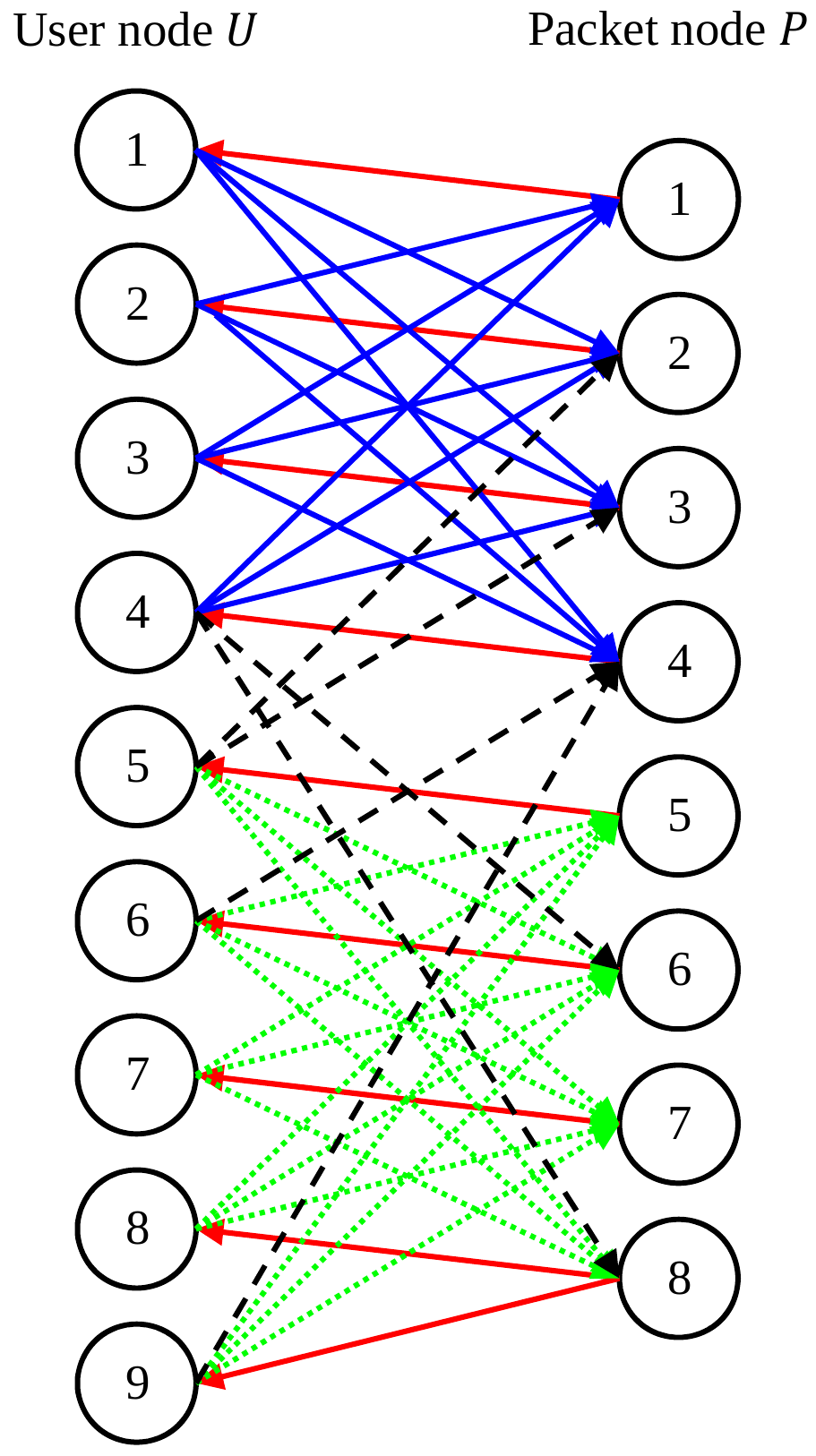}
\label{fig:fig7}}
\subfigure[]{\includegraphics[scale=0.5]{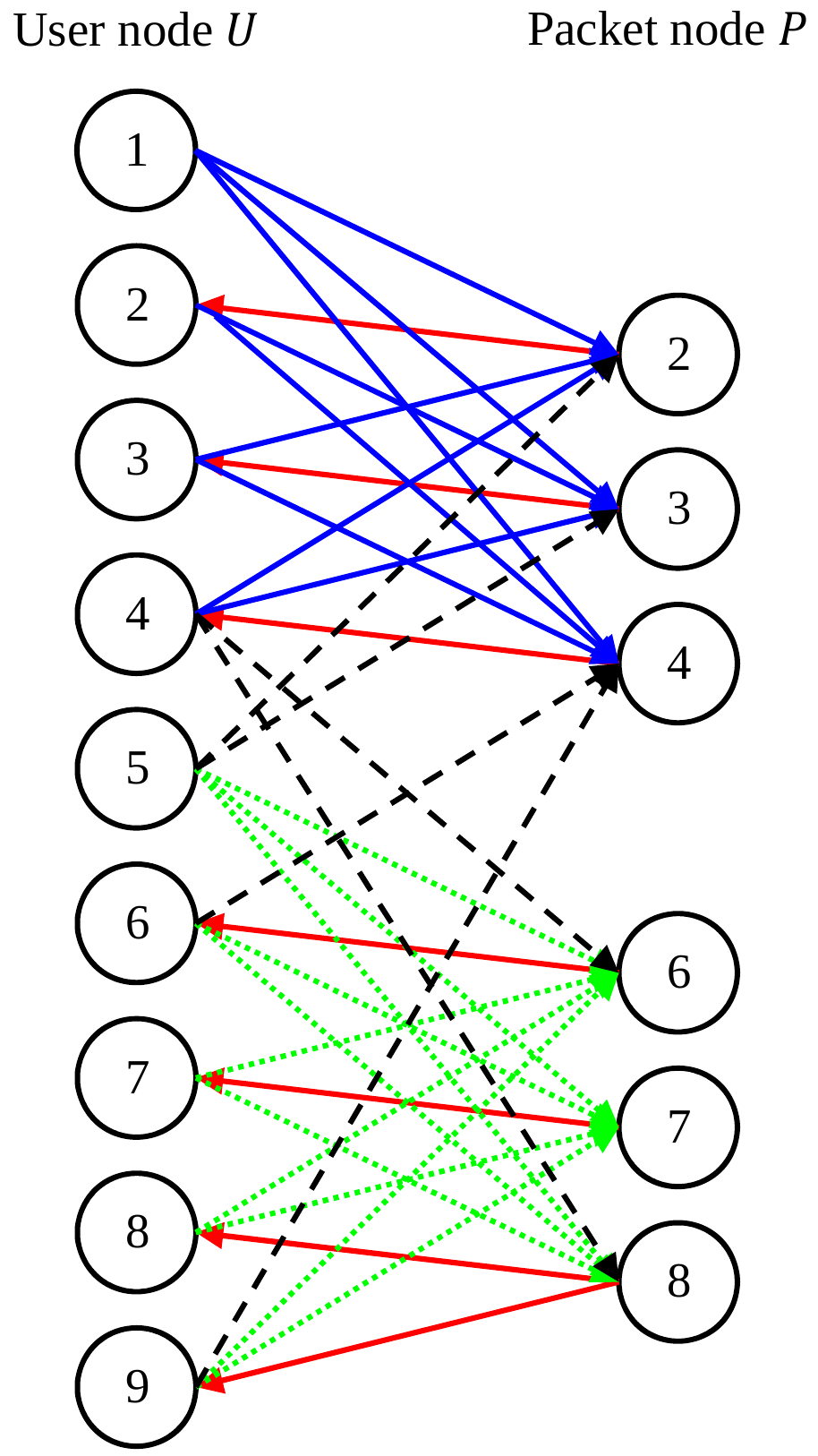}
\label{fig:fig8}}
\caption{The side information graphs of Example \ref{ex:cor}: (a) The side information graph $\mathcal{G}$. (b) The side information graph $\mathcal{G}^{\prime}$ after removing packet nodes $1$ and $5$.} 
\end{figure}

\begin{IEEEproof}
It is trivial that $n-\beta$ can be an upper bound for $N_{\rm opt}^q(\delta_s,\mathcal{G})$ from the above property 2) and Theorem \ref{theorem:delta cycle}. Thus, it is enough to show that $n-\beta$ can be a lower bound. If we remove one packet node and the corresponding edges from each $\delta_s$-cycle, the optimal codelength for the resulting graph is given as $N_{\rm opt}^q(\delta_s,\mathcal{G}^{\prime})=n-\beta$ because $\mathcal{G}^{\prime}$ is $\delta_s$-acyclic. Also, an index code for $\mathcal{G}$ satisfies the property of the generator matrix for ${\mathcal{G}}^{\prime}$ if we just substitute values of the removed packet nodes with $0$. Thus, we can say $N_{\rm opt}^q(\delta_s,\mathcal{G}^{\prime})\leq N_{\rm opt}^q(\delta_s,\mathcal{G})$. That is, $n-\beta=N_{\rm opt}^q(\delta_s,\mathcal{G})$.
\end{IEEEproof}
\end{cor}

\vspace{2mm}

An example for the previous corollary is provided.
\vspace{2mm}
\begin{example}
Let $q=2, m=9, n=8$, and $\delta_s=1$. A side information graph $\mathcal{G}$ is shown in Fig. \ref{fig:fig7}, where the maximum number of disjoint $\delta_s$-cycles is $2$. We denote two $\delta_s$-cycle induced sets as $B_1$ and $B_2$, that is, $B_1=\{1,2,3,4\}$ and $B_2=\{5,6,7,8\}$. The edges for $B_1$ and $B_2$ are represented by solid and dotted lines, respectively. If we remove the packet nodes $1$ and $5$ and the corresponding edges from two $\delta_s$-cycles, there is no $\delta_s$-cycle in the resulting side information graph as shown in Fig. \ref{fig:fig8}. Then, the optimal codelength is $6$ from Corollary \ref{cor:delta cycle}.
\label{ex:cor}
\end{example}
\vspace{2mm}
\begin{remark}
When $|\mathcal{X}_i|=2\delta_s+1$ for all $i\in Z[m]$, we have another example for Corollary \ref{cor:delta cycle}.
\end{remark}

\subsection{Clique}
We consider fully connected side information graphs in this section, that is, $|\mathcal{X}_i|=Z[n]\setminus f(i)$ for all $i\in Z[m]$. In the conventional index coding problem, the number of disjoint cliques is used as an upper bound of the optimal codelength and there are many heuristic algorithms to find the codelength based on the number of disjoint cliques. Therefore, we discuss cliques in the $(\delta_s,\mathcal{G})$-ICSIE problem. Since each receiver wants one symbol, we can consider the fully connected side information graphs as cliques. It is clear that cliques in index coding with erroneous side information are different from those of the conventional index coding, where cliques can be covered by one transmission for any fields. 

A set of vectors is said to be $(2\delta_s+1)$-linearly independent if any $2\delta_s+1$ vectors in the vector set are linearly independent. Since all vectors whose Hamming weight is less than or equal to $2\delta_s+1$ belong to $\mathcal{I}(q,\mathcal{G},\delta_s)$ when $\mathcal{G}$ is a clique, we have the following observation. 
\vspace{2mm}
\begin{observation}
In the $(\delta_s,\mathcal{G})$-ICSIE problem, finding the optimal codelength for cliques is equivalent to finding the minimum dimension of the set of $n$ vectors which is $(2\delta_s+1)$-linearly independent.
\label{Ob:clique}
\end{observation}
\vspace{2mm}

In general, $N_{\rm opt}^q(\delta_s,\mathcal{G})$ goes to $2\delta_s+1$ as the size of the finite field goes to infinity. If the size of a clique is less than or equal to $2\delta_s+1$, it is easy to check that $N_{\rm opt}^q(\delta_s,\mathcal{G})=n$. Thus, we consider the clique of size $n>2\delta_s+1$.
\vspace{2mm}
\begin{theorem}
There are some special cases of cliques for the $(\delta_s,\mathcal{G})$-ICSIE whose optimal codelength can be found as:
\begin{enumerate}
\item When $n=2\delta_s+2$, $N_{\rm opt}^q(\delta_s,\mathcal{G})=2\delta_s+1$ over $\mathbb{F}_q$.
\item When $\delta_s=1$, $N_{\rm opt}^q(\delta_s,\mathcal{G})$ is the minimum value of $N$ satisfying $2^{N-1}\geq n$ over $\mathbb{F}_2$.
\item If there are $N$ and $n$ satisfying $N\leq{{q+1}\over{q}}(2\delta_s+1)-1$ and $n\leq N+1$ over $\mathbb{F}_q$, $N_{\rm opt}^q(\delta_s,\mathcal{G})$ is the minimum value of $N$.
\item When $n=3\delta_s+3$, $N_{\rm opt}^q(\delta_s,\mathcal{G})=3\delta_s+1$ over $\mathbb{F}_2$.
\end{enumerate}
\vspace{2mm}
\begin{IEEEproof}

\begin{enumerate}
\item It is directly proved from Corollary \ref{cor:delta cycle} because $|\mathcal{X}_i|=2\delta_s+1$ for all $i\in Z[m]$. 
\item Let $Ind_q(N,k)$ be the maximal cardinality of the $k$-linearly independent subset of $\mathbb{F}_q^N$. We have $Ind_2(N,3)=2^{N-1}$ from \cite{NLI}, where $N\geq3$. Thus, we can think $N$ as a codelength and $k=2\delta_s+1$. To achieve the codelength $N$, the number of messages should be less than or equal to $Ind_q(N,k)$.  
\item From Theorem 2 in \cite{CLI}, for $2\leq k\leq N$, we have $Ind_q(N,k)=N+1$ if and only if ${{q}\over{q+1}}(N+1)\leq k$.
\item From \cite{NLI}, we have $Ind_2(N, N-m)=N+2$, where $N=3m+i, i=0,1$, and $m\geq2$. Since $N-m=2\delta_s+1$, we have $i=1$ and $m=\delta_s$. Then, for $N=3\delta_s+1$ and $\delta_s\geq2$, we have $Ind_2(N,2\delta_s+1)=N+2$. If $\delta_s=1$, $N_{\rm opt}^q(\delta_s,\mathcal{G})=3\delta_s+1$ by 2) when $n=3\delta_s+3$.
\end{enumerate}
\end{IEEEproof}
\label{theorem:clique case}
\end{theorem}

\vspace{2mm}
\begin{remark}
Construction of a generator matrix of each case in Theorem \ref{theorem:clique case} is also well defined. For example, a matrix whose rows consist of any vectors having odd weight in $\mathbb{F}_2^{N}$ can be a generator matrix in the above case 2).
\end{remark}
\vspace{2mm}

Next, we show how to construct a generator matrix of the $(\delta_s,\mathcal{G})$-ICSIE from a parity check matrix $H$ of an $(n,k,d_{min})$ classical error correcting code in the following proposition.

\vspace{2mm}
\begin{proposition}
Let $\mathcal{G}$ be the clique of size $n$ and $\bar{H}$ be the matrix having the smallest $n-k$ among $H$ of $(n,k,d_{min})$ classical error correcting codes with $d_{min}\geq2\delta_s+2$. Then, $\bar{H}^{\intercal}$ becomes the optimal generator matrix of the $(\delta_s,\mathcal{G})$-ICSIE.
\begin{figure}[t]
\centering
\includegraphics[scale=0.5]{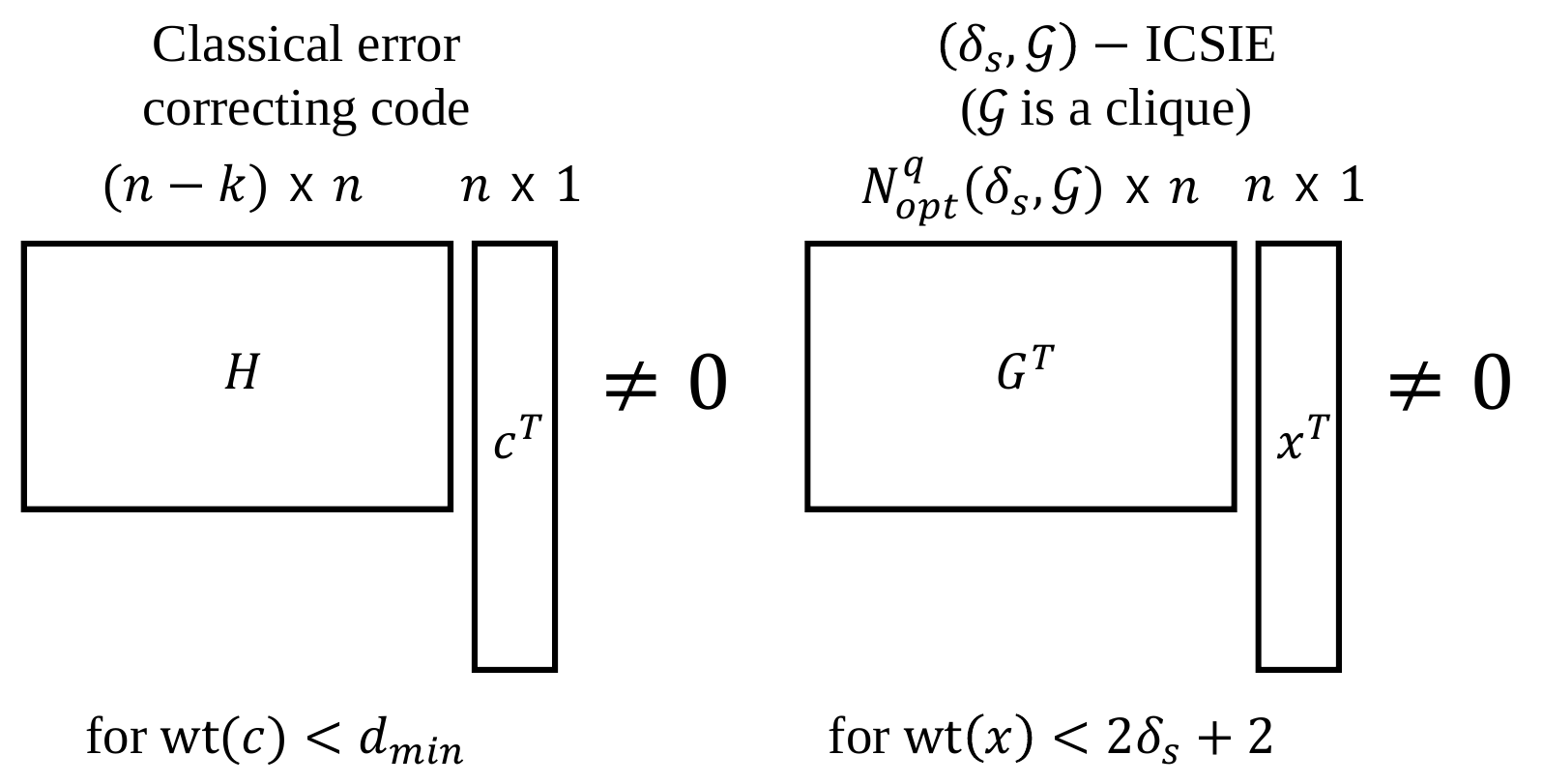}
\caption{Comparison between the classical error correcting code and the $(\delta_s,\mathcal{G})$-ICSIE.} 
\label{fig:fig9}
\end{figure}
\vspace{2mm}
\begin{IEEEproof}
In the $(\delta_s,\mathcal{G})$-ICSIE problem, where a side information graph $\mathcal{G}$ is the  clique of size $n$, it is clear that any $2\delta_s+1$ row vectors of a generator matrix are linearly independent by Observation \ref{Ob:clique}. That is, $\bold{x}G\neq\bold{0}$ for any $\bold{x}$ such that ${\rm wt}(\bold{x})\leq2\delta_s+1$. Then, we can easily check that an $n\times (n-k)$ matrix $H^{\intercal}$ can be a generator matrix of a $(\delta_s,\mathcal{G})$-ICSIE if $d_{min}\geq2\delta_s+2$ as shown in Fig. \ref{fig:fig9}. Thus, when a side information graph $\mathcal{G}$ is the clique of size $n$, the optimal codelength of the $(\delta_s,\mathcal{G})$-ICSIE is the minimum value of $n-k$ for an $(n,k,d_{min})$ classical error correcting code satisfying $d_{min}\geq2\delta_s+2$.
\end{IEEEproof}
\label{prop:parity}
\end{proposition}
\vspace{2mm}
\begin{remark}
One of examples is the Reed-Solomon code when $n$ divides $q-1$. In this case, the optimal codelength of the $(\delta_s,\mathcal{G})$-ICSIE is $n-k=d_{min}-1=2\delta_s+1$.
\end{remark}
\vspace{2mm}
\begin{remark}
Even if $\mathcal{G}$ is not a clique, we can regard the parity check matrix of the classical error correcting code as the transpose of the generator matrix of the $(\delta_s,\mathcal{G})$-ICSIE when $d_{min}$ of the error correcting code is larger than the maximum weight of vectors in $\mathcal{I}(q,\mathcal{G},\delta_s)$. 
\end{remark}
\subsection{Lower Bounds for the Optimal Codelength}

It is not difficult to check the following corollary and observations for the $(\delta_s,\mathcal{G})$-ICSIE.
\vspace{2mm}
\begin{cor}
Let $S=\{j\in Z[n]|\exists i\in Z[m] \textrm{ s.t. } f(i)=j \textrm{ and } |\mathcal{X}_i|\leq2\delta_s\}$. Then, we have $N_{\rm opt}^q(\delta_s,\mathcal{G})\geq|S|+1$ for $n>|S|$.
\begin{IEEEproof}
It is clear that $\mathcal{I}_i(q,\mathcal{G},\delta_s)=\{\bold{z}\in \mathbb{F}_q^n: z_{f(i)}\neq 0\}$, that is, $G_{f(i)}$ does not belong to span$(\{G_j\}_{j\in Z[n]\setminus f(i)})$ for $f(i)\in S$. Thus, the corollary is obvious.
\end{IEEEproof}
\end{cor}
\vspace{2mm}
\begin{remark}
Thus, having less than or equal to $2\delta_s$ side information symbols is the same as not having side information in index coding with erroneous side information.
\end{remark}
\vspace{2mm}
\begin{observation}
For the given $(\delta_s,\mathcal{G})$-ICSIE problem, let $\mathcal{G}^{\prime}$ be an edge-induced subgraph obtained by deleting some outgoing edges of user nodes of $\mathcal{G}$. Then, $N_{\rm opt}^q(\delta_s,\mathcal{G})\leq N_{\rm opt}^q(\delta_s,\mathcal{G}^{\prime})$.
\label{ob:subgraph}
\end{observation}
\vspace{-20pt}
\vspace{2mm}
\begin{observation}
If $\delta_s^{\prime}\leq\delta_s$,  $N_{\rm opt}^q(\delta_s^{\prime},\mathcal{G})\leq N_{\rm opt}^q(\delta_s,\mathcal{G})$.
\label{ob:delta}
\end{observation}
\vspace{2mm}

Now, the relationship of the optimal codelength between the conventional index code and the proposed ICSIE is given in the following theorem.

\vspace{2mm}
\begin{theorem}
Suppose that the $(0,\bar{\mathcal{G}})$-IC problem is constructed by deleting any min$(2\delta_s, |\mathcal{X}_i|)$ outgoing edges from each receiver $R_i$ in the $(\delta_s,\mathcal{G})$-ICSIE problem. That is, each receiver of $\bar{\mathcal{G}}$ has max$(0, |\mathcal{X}_i|-2\delta_s)$ side information symbols and then it becomes the conventional index coding problem. Then, $N_{\rm opt}^q(0,\bar{\mathcal{G}})\leq N_{\rm opt}^q(\delta_s,\mathcal{G})$.
\vspace{2mm}
\begin{IEEEproof}
A generator matrix of the $(\delta_s,\mathcal{G})$-ICSIE problem can be a generator matrix of the $(0,\bar{\mathcal{G}})$-IC problem because $\mathcal{I}(q,\bar{\mathcal{G}},0)\subseteq\mathcal{I}(q,\mathcal{G},\delta_s)$. Specifically, for a vector $\bold{z}^{\prime}\in\mathcal{I}_i(q,\bar{\mathcal{G}},0)$, ${\rm wt}(\bold{z}^{\prime}_{\mathcal{X}_i})\leq2\delta_s$ since ${\rm wt}(\bold{z}^{\prime}_{\mathcal{X}^{\prime}_i})$ should be zero, where $\mathcal{X}^{\prime}_i$ is a set of side information indices of $R_i$ for $\bar{\mathcal{G}}$. Thus, $\mathcal{I}_i(q,\bar{\mathcal{G}},0)\subseteq\mathcal{I}_i(q,\mathcal{G},\delta_s)$. Since it is true for all $i\in Z[m]$, $\mathcal{I}(q,\bar{\mathcal{G}},0)\subseteq\mathcal{I}(q,\mathcal{G},\delta_s)$.
\end{IEEEproof}
\label{theorem:lower bound}
\end{theorem}
\vspace{2mm}
\begin{remark}
In general, if we reduce $\delta_s$, we can have a lower bound from Observation \ref{ob:delta}. Similarly, if we delete outgoing edges of user nodes, we can have an upper bound from Observation \ref{ob:subgraph}. However, if we reduce $\delta_s$ to $0$ and delete min$(2\delta_s, |\mathcal{X}_i|)$ outgoing edges of each receiver $R_i$, we can have a lower bound as in Theorem \ref{theorem:lower bound}. Thus, the worst case of the resulting $(0,\bar{\mathcal{G}})$-IC problems can be a lower bound for the corresponding $(\delta_s,\mathcal{G})$-ICSIE problem.
\end{remark}
\vspace{2mm}
\begin{example}
Let $q=2, n=m=4, \delta_s=1, f(i)=i$, and $\mathcal{G}$ as shown in Fig. \ref{fig:fig4}. Then, we have $N_{\rm opt}^q(1,\mathcal{G})=4$.
\begin{figure}[t]
\centering
\subfigure[]{\includegraphics[scale=0.72]{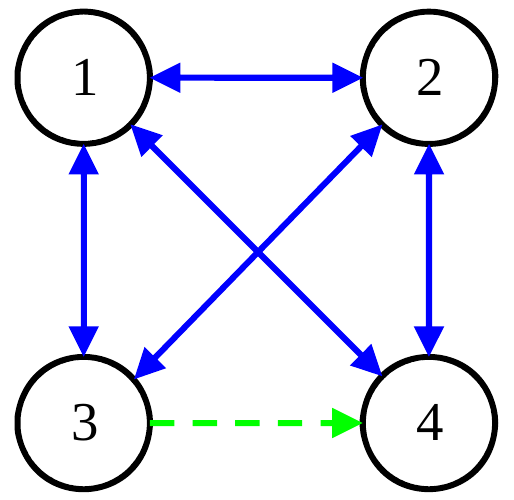}
\label{fig:fig4}}
\subfigure[]{\includegraphics[scale=0.72]{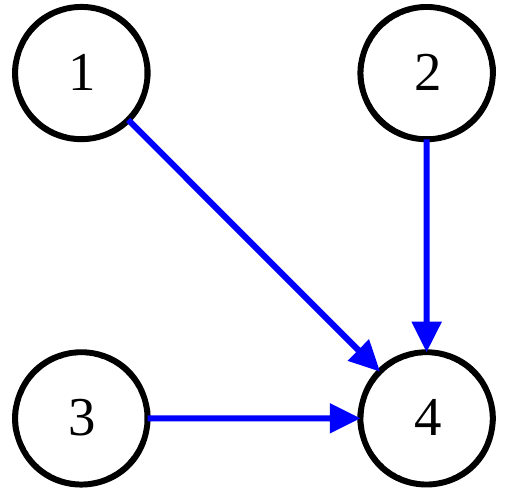}
\label{fig:fig5}}
\caption{The unipartite side information graphs of Example \ref{ex:noclique}: (a) The unipartite side information graph $\mathcal{G}$. (b) The unipartite side information graph $\bar{\mathcal{G}}$.} 
\label{fig:example 4}
\end{figure} If we delete 2 outgoing edges from each receiver, there is a side information graph $\bar{\mathcal{G}}$ as shown in Fig. \ref{fig:fig5}. In the conventional index coding problem, $N_{\rm opt}^q(0,\bar{\mathcal{G}})=n$ if a side information graph $\bar{\mathcal{G}}$ is acyclic \cite{DIC}. Since the graph in Fig. \ref{fig:fig5} is acyclic, $N_{\rm opt}^q(0,\bar{\mathcal{G}})=4$. From Theorem \ref{theorem:lower bound}, $N_{\rm opt}^q(1,\mathcal{G})=4$ because $4=N_{\rm opt}^q(0,\bar{\mathcal{G}})\leq N_{\rm opt}^q(1,\mathcal{G})\leq4=n$.
\label{ex:noclique}
\end{example}
\vspace{2mm}
\begin{example}
Let $q=2, n=m=4, \delta_s=1, f(i)=i$, and $\mathcal{G}$ is the clique of size 4 as in Example \ref{ex: encoding}. If we delete 2 outgoing edges from each receiver, the corresponding graph has at least one cycle because each receiver has one outgoing edge. In this case, we can reduce the codelength by at least one because all cycles in the graph consist of unicast packets \cite{DIC}. Then, the worst case of the corresponding graph $\bar{\mathcal{G}}$ has $N_{\rm opt}^q(0,\bar{\mathcal{G}})=3$. Thus, $3\leq N_{\rm opt}^q(1,\mathcal{G})\leq4$. In fact, we have $3\leq N_{\rm opt}^q(1,\mathcal{G})=3$ because there is a generator matrix of the $(\delta_s,\mathcal{G})$-ICSIE given by 
\begin{equation}
G=\begin{pmatrix}1&0&0\\0&1&0\\0&0&1\\1&1&1\end{pmatrix}.
\nonumber
\end{equation}
\end{example}
\vspace{2mm}

It is easy to derive the following lower bound for the optimal codelength.
\vspace{2mm}
\begin{theorem}
$N_{\rm opt}^q(\delta_s,\mathcal{G})\geq\gamma(\mathcal{G})$.
\vspace{2mm}
\begin{IEEEproof}
From the definition of $\gamma(\mathcal{G})$, the corresponding $\gamma(\mathcal{G})$ rows of a generator matrix of the $(\delta_s,\mathcal{G})$-ICSIE should be linearly independent.
\end{IEEEproof}
\end{theorem}
\vspace{2mm}

\vspace{10pt}
\section{Generalized Error Correcting Index Codes}\label{sec:Generalization}

We can generalize many properties of the ECIC in \cite{SECIC} by considering the side information errors. That is, we can describe the properties of the $(\delta_s,\delta_c,\mathcal{G})$-GECIC similar to those of the $(\delta_c,\mathcal{G})$-ECIC in \cite{SECIC} by using the properties of the $(\delta_s,\mathcal{G})$-ICSIE. Some notations of the ECIC in \cite{SECIC} are changed for consistency within paper.
\vspace{2mm}
\begin{proposition}
Properties of the $(\delta_c,\mathcal{G})$-ECIC in \cite{SECIC} can be generalized to those of the $(\delta_s,\delta_c,\mathcal{G})$-GECIC as:

\begin{enumerate}
\item Generalization of Lemma 3.8 in \cite{SECIC};

Theorem \ref{theorem:Encoding}
\item Generalization of Proposition 4.6 in \cite{SECIC};

$N_{\rm opt}^q(\delta_s,\delta_c,\mathcal{G})\leq l_q[N_{\rm opt}^q(\delta_s,\mathcal{G}),2\delta_c+1]$, where $l_q[a,b]$ denotes the minimum codelength for the dimension $a$ and $d_{min}=b$.
\item Generalization of Theorem 5.1 in \cite{SECIC};

$N_{\rm opt}^q(\delta_s,\delta_c,\mathcal{G})\geq N_{\rm opt}^q(\delta_s,\mathcal{G})+2\delta_c$.
\item Generalization of the property of $\gamma({\mathcal{G}})$ in \cite{SECIC};

When $m=n$ and $f(i)=i$ for all $i\in Z[n]$, $\gamma({\mathcal{G}})=\delta_s$-MAIS$(\mathcal{G})$, where $\delta_s$-MAIS$(\mathcal{G})$ denotes the maximum size of a $\delta_s$-acyclic induced subgraph of $\mathcal{G}$.
\item Generalization of Theorem 4.3 in \cite{SECIC};

$N_{\rm opt}^q(\delta_s,\delta_c,\mathcal{G})\geq l_q[\gamma(\mathcal{G}),2\delta_c+1]$.
\end{enumerate}
\vspace{2mm}
\begin{IEEEproof}
All generalization except 4) can be easily proved by the same methods as in \cite{SECIC} if we replace the conventional index code with the $(\delta_s,\mathcal{G})$-ICSIE. In the case of 4), we already prove an equivalence between a $\delta_s$-generalized independent set and a $\delta_s$-acyclic graph in Lemma \ref{lemma:equivalence}.
\end{IEEEproof}
\label{prop:general}
\end{proposition}
\vspace{2mm}
\begin{remark}
By 4) of Proposition \ref{prop:general}, we can think that a $\delta_s$-cycle corresponds to a cycle in  the conventional index coding.
\end{remark}

\vspace{2mm}

Now, we introduce some properties of the $(\delta_s,\delta_c,\mathcal{G})$-GECIC.
\vspace{2mm}
\begin{theorem} 
Suppose that the $(\delta_c,\bar{\mathcal{G}})$-ECIC problem is constructed by deleting any min$(2\delta_s, |\mathcal{X}_i|)$ outgoing edges from each receiver $R_i$ in the $(\delta_s,\delta_c,\mathcal{G})$-GECIC problem. That is, each receiver of $\bar{\mathcal{G}}$ has max$(0, |\mathcal{X}_i|-2\delta_s)$ side information symbols and then it becomes the conventional error correcting index coding problem. Then, $N_{\rm opt}^q(\delta_c,\bar{\mathcal{G}})\leq N_{\rm opt}^q(\delta_s,\delta_c,\mathcal{G})$.

\begin{figure}[t]
\centering
\includegraphics[scale=0.5]{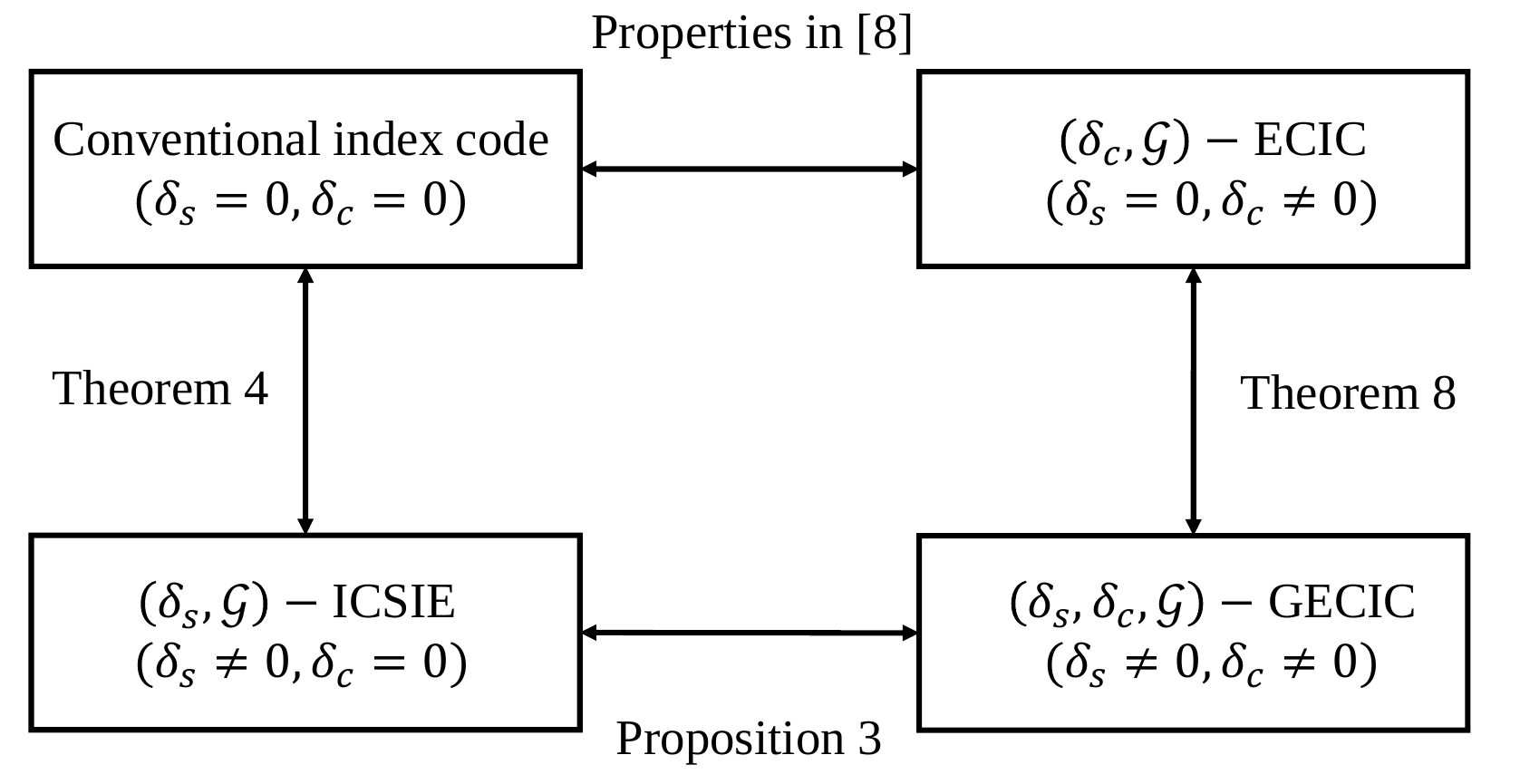}
\caption{Relationship of several index coding problems for the optimal codelength.} 
\label{fig:fig10}
\end{figure}

\vspace{2mm}
\begin{IEEEproof}
The proof is similar to that of Theorem \ref{theorem:lower bound} and thus we omit it.
\end{IEEEproof}
\end{theorem}
\vspace{2mm}
\begin{theorem}
Let $\hat{\mathcal{G}}$ be an edge-induced subgraph of $\mathcal{G}$, which is obtained by deleting all outgoing edges of all users in $\mathcal{G}$. Then, $N_{\rm opt}^q(\delta_s,\delta_c,\mathcal{G})=N_{\rm opt}^q(\delta_s,\delta_c,\hat{\mathcal{G}})$ if $\Phi=\phi$.
\begin{IEEEproof}
If $\Phi=\phi$, $Z[n]$ is a $\delta_s$-generalized independent set. Then, $\mathcal{I}(q,\mathcal{G},\delta_s)=\mathcal{I}(q,\hat{\mathcal{G}},\delta_s)$ and thus $N_{\rm opt}^q(\delta_s,\delta_c,\mathcal{G})=N_{\rm opt}^q(\delta_s,\delta_c,\hat{\mathcal{G}})$.
\end{IEEEproof}
\end{theorem}

\vspace{2mm}

In Fig. \ref{fig:fig10}, we show the relationship between the proposed index codes and several index coding problems, specifically in terms of the optimal codelength. 
\vspace{10pt}
\section{Conclusions}\label{Sec:Conclusion}
We generalized the index coding problem, where there is a possibility to have erroneous side information in each receiver. The property of the generator matrix and the decoding procedure of the proposed index codes with erroneous side information were suggested, which are based on the idea of Hamming spheres and the syndrome decoding, respectively. 

We also suggested some bounds for the optimal codelength of the $(\delta_s,\mathcal{G})$-ICSIE and showed the relationship between the conventional index coding and index coding with erroneous side information. In addition, we found a $\delta_s$-cycle of the GECIC, which has similar properties with those of a cycle in the conventional index coding. It was also found that the existence of a $\delta_s$-cycle is crucial in the proposed index coding problem. The proposed ICSIE was also analyzed when a side information graph $\mathcal{G}$ is a clique. Through this, it was found that the generator matrix for the $(\delta_s,\mathcal{G})$-ICSIE corresponds to the transpose of the parity check matrix of the classical error correcting code when the related parameters are properly chosen. 

Finally, it was shown that the existing bounds and properties for the $(\delta_c,\mathcal{G})$-ECIC can be generalized to those of the $(\delta_s,\delta_c,\mathcal{G})$-GECIC by using the properties of the $(\delta_s,\mathcal{G})$-ICSIE. That is, we can easily derive bounds for the optimal codelength of the $(\delta_s,\delta_c,\mathcal{G})$-GECIC, which is the index code in the more generalized scenario.

\end{document}